\pdfoutput=1

\documentclass[sigconf]{acmart}

\usepackage{symbols}
\usepackage{bm}
\usepackage{listings}
\usepackage{multirow}
\usepackage[utf8]{inputenc}
\usepackage[shortlabels]{enumitem}
\setlist[enumerate]{leftmargin=5mm,topsep=1mm}
\setlist[itemize]{leftmargin=5mm,topsep=1mm}
\newcommand{\capture}{\textsf{Capture}\xspace}

\newcommand{\ignore}[1]{}

\definecolor{Acolor}{HTML}{264653}
\definecolor{Bcolor}{HTML}{2A9D8F}
\definecolor{Ccolor}{HTML}{E9C46A}
\definecolor{Dcolor}{HTML}{F4A261}
\definecolor{Ecolor}{HTML}{E76F51}

\AtBeginDocument{%
  \providecommand\BibTeX{{%
    \normalfont B\kern-0.5em{\scshape i\kern-0.25em b}\kern-0.8em\TeX}}}

\copyrightyear{2023} 
\acmYear{2023} 
\setcopyright{acmlicensed}\acmConference[DEEM '23]{Data Management for End-to-End Machine Learning }{June 18, 2023}{Seattle, WA, USA}
\acmBooktitle{Data Management for End-to-End Machine Learning (DEEM '23), June 18, 2023, Seattle, WA, USA}
\acmPrice{15.00}
\acmDOI{10.1145/3595360.3595855}
\acmISBN{979-8-4007-0204-4/23/06}

\begin{document}

\title[Transactional Python for Durable Machine Learning]{Transactional Python for Durable Machine Learning:
\\ Vision, Challenges, and Feasibility
}

\author{Supawit Chockchowwat,
Zhaoheng Li,
Yongjoo Park}
\email{{supawit2,zl20,yongjoo}@illinois.edu}
\affiliation{%
  \country{University of Illinois at Urbana-Champaign}
}

\renewcommand{\shortauthors}{Chockchowwat, Li, and Park}

\begin{abstract}
    In machine learning (ML),
        Python serves as a convenient abstraction for working with 
    key libraries such as PyTorch, scikit-learn, and others.
    Unlike DBMS, however, \textit{Python applications may lose important data}, 
        such as trained models and extracted features, 
    due to machine failures or human errors,
        leading to a waste of time and resources.
    Specifically, they lack four essential properties that could make ML more reliable and 
        user-friendly---\textit{durability, atomicity, replicability, and time-versioning (DART)}.

    This paper presents our vision of \emph{Transactional Python
        that provides DART} without any code modifications to user programs or the Python kernel,
    by non-intrusively monitoring application states at the object level
        and determining a minimal amount of information
            sufficient to reconstruct a whole application.
    Our evaluation of a proof-of-concept implementation with public PyTorch and scikit-learn applications
        shows that DART can be offered with overheads ranging 1.5\%--15.6\%.
\end{abstract}

\maketitle

\section{Introduction}
\label{sec:intro}

From interactive AI to scientific computing,
Python is widely used\footnote{There are 15 million active 
    Python developers worldwide as of Q1 2022~\cite{slashdata}.} 
for bringing together key tools
    such as 
TensorFlow~\cite{abadi2016tensorflow}, PyTorch~\cite{paszke2019pytorch}, 
    Pandas~\cite{mckinney2011pandas}, and Modin~\cite{petersohn2020towards}.
\ignore{
As one of the most popular programming languages~\cite{??}, Python has attracted programmers from diverse backgrounds.
Most importantly, its extensible designs have allowed these diverse programmers to contribute from their interests, inducing abstraction layers that hide implementation details in performant languages while exposing higher-level interfaces. }
As an interpreted language,
Python statements often serve as abstraction layers 
    hiding sophisticated implementations written in lower-level languages (e.g., C).
While its syntax differs from SQL,
the fundamental nature of their operations remains the same:
in both Python and SQL,
    statements may alter the state of data managed by the application or the system.

Unlike DBMS (e.g., PostgreSQL), however,
    \textit{Python applications are prone to data loss};
at any point, 
they may unexpectedly terminate  due to human errors or engineering issues,
    losing all of its intermediate data.
That is, 
\textit{Python applications
    lack durability, atomicity, replicability, and time-versioning}.
\emph{No Durability:} if an app terminates, no state persists.
\emph{No Atomicity:} 
    an interrupted function may result in partial updates.
\emph{No Easy Replication:}
    the state of a running app cannot be easily
    replicated to another machine and resumed there.
\emph{No Versioning:} 
    we cannot pick up a past state
        (e.g., models in a previous epoch).
While orchestration services~\cite{k8s,naik2016building}
    can enhance system robustness,
        they cannot restore data.
Manual \emph{point solutions}~\cite{zaharia2018accelerating, vartak2016modeldb, python-object-store}
    may miss important information,
    are time-consuming,
    and incur unnecessary computational costs.


\paragraph{Our Approach}

We aim to bring \emph{durability, atomicity, replicability, and time-versioning (DART)
    to Python}
by reinterpreting \emph{database transactions}
    and building a new framework.
DART is offered as follows.
\emph{Durability:} Application states will be automatically 
    persisted to survive unexpected errors or failures.
\emph{Atomicity:} 
    Only completed statements will result in
        valid states for persistence.
\emph{Replicability:} 
Running applications can be copied onto different machines and resumed.
\emph{Versioning:} Persisted states will allow users 
    to resume an application from a past state.
To accomplish DART,
    our work develops
efficient approaches to 
    identifying and expressing partial state changes made to an application state,
    without requiring changes to programs as well as
        the Python kernel (\cref{fig:overall}).

\ignore{
At the core, this is a data management problem with Python state as the data. That is, we need state persistence to enable solutions such as low-effort model checkpointing, time travel model diagnosis, and fault recovery (\cref{sec:overview}). Our solution must quietly function under desired overhead allowances (e.g., time, memory, I/O, and critical sections). Furthermore, it must not require any code modification to function properly by default. With a proper interface to database management systems (DBMS) evolved to handle demanding workloads, we can alleviate this problem and let Python programmers focus on their abstraction layer once again.
}

\ignore{
To do so, we propose a concept to view Python as a transaction, bridging Python setting to data management techniques (\cref{sec:tpy}). Then, we lay out a system sketch, listing essential components to realize the solution (\cref{sec:system}). Our preliminary feasibility study shows that, based on the concept and sketch, our proof-of-concept implementation can capture various states of Python-for-ML applications over time at a reasonable overhead (\cref{sec:exp}). Finally, we conclude with our ongoing works to tackle open data management problems.
}

\textit{Example Scenario to be Addressed:}
To see the relationship between childcare types and developments,
    Jane conducts regression analyses using shared computing resources
        managed via Slurm.
Jane submits her job with 4 hrs of time allocation.
After a while,
    Jane checks the output in the log:
her job completed the regression, but 
    terminated almost at the last moment 
    due to permission issues
    while saving the results to a file.
Now, Jane needs to start from scratch 
    by re-submitting the job after fixing the issue.
The same problem can happen
    when the job terminates after exceeding time limits,
or as the shared system enters into brief maintenance.
\textit{Our work will
enable a new service
    allowing Jane to load and resume from  any past application state 
        effortlessly.}

\paragraph{Comparison to Existing Work}
Our approaches offer technical advantages over existing methods in managing data-intensive applications~\cite{sheoran2023stepdeepola,li2023sc}. 
For instance, memory snapshots can be created at an OS level to retain past application states like CRIU~\cite{venkatesh2019fast}.
However, this method incurs prohibitive costs with data redundancy and is platform-dependent.
Alternatively, programmers can manually record data using external services like MLFlow~\cite{zaharia2018accelerating}, Mistique~\cite{vartak2018mistique}, and Ambrosia~\cite{goldstein2020ambrosia},
but this method requires extensive code modifications, 
may not capture all necessary information, 
and cannot leverage program structure and intra-data dependencies
    for optimization.
DBOS~\cite{skiadopoulos2021DBOS} brings DBMS capabilities to serve OS services like file systems, scheduler, and inter-process communication; however, like other OS's, DBOS's services do not include taking a snapshot of application states. To do so, DBOS needs additional techniques like our work.
Please see \cref{tab:intro-related} for more related work.

\begin{table}[t]

\centering

\caption{Limitations of Existing Approaches.}
\label{tab:intro-related}
\vspace{-4mm}
\footnotesize
\renewcommand{\arraystretch}{1.1}
\begin{tabular}{ l l l }
\toprule
 \textbf{Approach} & \textbf{Limitation} \\ 
\midrule
Recompute All
    & Takes long if includes large-scale data operations \\
Time-Travel DB \cite{morrey2003peabody,soroush2013time,bhattacherjee2015principles}
    & Focuses on SQL-based data updates \\
Data Lineage \cite{phani2021lima, bose2005lineage, buneman2006provenance,
ruan2019fine}
    & Require efficient methods for identifying changes \\
Reverse Debugging \cite{zelkowitz1973reversible, feldman1988igor}
    & Too large overhead for regular applications \\
OS-level Snapshot \cite{juric2021checkpoint, jain2020crac}
    & Efficient due to high data redundancy \\
Save All Objects \cite{dumpsession, jupyterstore}
    & Slow and incurs large storage costs \\
\bottomrule
\end{tabular}

\vspace{-4mm}

\end{table}

\ignore{
\paragraph{Contributions}
We first describe our system design for transaction Python (\cref{sec:system}),
and how we solve a core technical challenge
    for efficiently storing application states (\cref{sec:state}).
We also study its feasibility empirically (\cref{sec:exp}).
Finally, we discuss 
    remaining challenges (\cref{sec:conclusion}).
}

\section{Transactional Python System}
\label{sec:overview}
\label{sec:system}



\begin{figure}[t]
  \centering
  \tikzset{every picture/.style={line width=0.75pt}} 

\begin{tikzpicture}[x=0.75pt,y=0.75pt,yscale=-1,xscale=1]

\draw   (40,30) -- (110,30) -- (110,60) -- (40,60) -- cycle ;
\draw  [fill={rgb, 255:red, 222; green, 235; blue, 252 }  ,fill opacity=1 ] (40,80) -- (110,80) -- (110,100) -- (40,100) -- cycle ;
\draw    (75,60) -- (75,78) ;
\draw [shift={(75,80)}, rotate = 270] [color={rgb, 255:red, 0; green, 0; blue, 0 }  ][line width=0.75]    (10.93,-3.29) .. controls (6.95,-1.4) and (3.31,-0.3) .. (0,0) .. controls (3.31,0.3) and (6.95,1.4) .. (10.93,3.29)   ;
\draw   (30,20) -- (120,20) -- (120,110) -- (30,110) -- cycle ;
\draw  [fill={rgb, 255:red, 255; green, 255; blue, 255 }  ,fill opacity=1 ] (170,94.24) -- (170,102.72) .. controls (170,103.98) and (163.28,105) .. (155,105) .. controls (146.72,105) and (140,103.98) .. (140,102.72) -- (140,94.24)(170,94.24) .. controls (170,95.5) and (163.28,96.52) .. (155,96.52) .. controls (146.72,96.52) and (140,95.5) .. (140,94.24) .. controls (140,92.98) and (146.72,91.96) .. (155,91.96) .. controls (163.28,91.96) and (170,92.98) .. (170,94.24) -- cycle ;
\draw    (110,90) -- (138,90) ;
\draw [shift={(140,90)}, rotate = 180] [color={rgb, 255:red, 0; green, 0; blue, 0 }  ][line width=0.75]    (10.93,-3.29) .. controls (6.95,-1.4) and (3.31,-0.3) .. (0,0) .. controls (3.31,0.3) and (6.95,1.4) .. (10.93,3.29)   ;
\draw  [fill={rgb, 255:red, 222; green, 235; blue, 252 }  ,fill opacity=1 ] (190,80) -- (260,80) -- (260,100) -- (190,100) -- cycle ;
\draw    (170,90) -- (188,90) ;
\draw [shift={(190,90)}, rotate = 180] [color={rgb, 255:red, 0; green, 0; blue, 0 }  ][line width=0.75]    (10.93,-3.29) .. controls (6.95,-1.4) and (3.31,-0.3) .. (0,0) .. controls (3.31,0.3) and (6.95,1.4) .. (10.93,3.29)   ;
\draw    (195,25) -- (195,80) ;
\draw    (195,25) -- (205,25) ;
\draw    (195,45) -- (205,45) ;
\draw    (195,65) -- (205,65) ;
\draw  [fill={rgb, 255:red, 255; green, 255; blue, 255 }  ,fill opacity=1 ] (170,85.76) -- (170,94.24) .. controls (170,95.5) and (163.28,96.52) .. (155,96.52) .. controls (146.72,96.52) and (140,95.5) .. (140,94.24) -- (140,85.76)(170,85.76) .. controls (170,87.02) and (163.28,88.04) .. (155,88.04) .. controls (146.72,88.04) and (140,87.02) .. (140,85.76) .. controls (140,84.5) and (146.72,83.48) .. (155,83.48) .. controls (163.28,83.48) and (170,84.5) .. (170,85.76) -- cycle ;
\draw  [fill={rgb, 255:red, 255; green, 255; blue, 255 }  ,fill opacity=1 ] (170,77.28) -- (170,85.76) .. controls (170,87.02) and (163.28,88.04) .. (155,88.04) .. controls (146.72,88.04) and (140,87.02) .. (140,85.76) -- (140,77.28)(170,77.28) .. controls (170,78.54) and (163.28,79.57) .. (155,79.57) .. controls (146.72,79.57) and (140,78.54) .. (140,77.28) .. controls (140,76.02) and (146.72,75) .. (155,75) .. controls (163.28,75) and (170,76.02) .. (170,77.28) -- cycle ;

\draw (75,90) node  [font=\footnotesize] [align=left] {Capture (\cref{sec:ui})};
\draw (75,45) node  [font=\footnotesize] [align=left] {User App.};
\draw (22.5,90) node  [font=\footnotesize,rotate=-270] [align=left] {Python};
\draw (225,90) node  [font=\footnotesize] [align=left] {Recovery (\cref{sec:overview:usecases})};
\draw (207,25) node [anchor=west] [inner sep=0.75pt]  [font=\footnotesize] [align=left] {Model Checkpoint};
\draw (207,45) node [anchor=west] [inner sep=0.75pt]  [font=\footnotesize] [align=left] {Time Travel Diagnosis};
\draw (207,65) node [anchor=west] [inner sep=0.75pt]  [font=\footnotesize] [align=left] {Fault Recovery};
\draw (155,72) node [anchor=south] [inner sep=0.75pt]  [font=\footnotesize] [align=left] {\begin{minipage}[lt]{26.3pt}\setlength\topsep{0pt}
\begin{center}
State\\DBMS
\end{center}

\end{minipage}};

\end{tikzpicture}
  \vspace{-3mm}
  \caption{Transactional Python's workflow overview.}
  \vspace{-4mm}
  \label{fig:overall}
\end{figure}
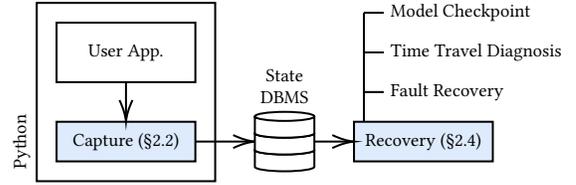

\ignore{
Our user interacts with the transactional Python through two additional modes: \emph{capture} and \emph{recovery} (\cref{fig:overall}). During the capture mode, the user first attaches our capture module to any Python application of their choice with no code modification required. This capture module collects the application's states quietly as the application runs normally. Afterwards during a recovery, the user has accesses to a wide variety of tools enabled by captured states such as low-effort model checkpointing, time-travel learning diagnosis, and automatic model restart. In a way, capture is analogous to write operations, while recovery is analogous to read operations.}

In this section,
    we formally state the problem (\cref{sec:overview:problem})
and describe how users can run their applications
    on our framework without modifications to the code 
        (\cref{sec:ui}).
We describe our framework's internal mechanism
    for monitoring a user application
to trace its states
    toward DART (\cref{sec:overview:dart}).
Our approach will enable new use cases (\cref{sec:overview:usecases})
    by overcoming technical challenges 
    (\cref{sec:overview:challenge}).

\subsection{High-Level Objectives}
\label{sec:overview:problem}

An application runs as an interpreter turns 
    a user program into statements, $M_1, \ldots, M_n$, where
        $M_{i+1}$ alters the state from $S_i$ to $S_{i+1}$,
yielding a series of states $S_1, \ldots, S_n$.
\textbf{\textit{Atomicity and Versioning:}}
$S_i$ is persisted after a \emph{complete} execution of $M_i$.
For an index $i \in {1, \ldots, n}$, our framework can roll back the application state to $S_i$,
    from which it may resume.
%
\textbf{\textit{Replication and Durability:}}
Our framework will enable efficient replication of $S_i$
    onto durable storage $D$ by storing
        information $R_i$ sufficient to reconstruct $S_i$ as quickly as possible.
That is, we will persist $\bm{R} = \{R_1, \ldots, R_n\}$ onto $D$
    to generate the entire application history $\bm{S} = \{S_1, \ldots, S_n\}$.
\textbf{\textit{State Composition:}}
$S_i$ consists of \emph{objects} 
    that may contain shared references;
for example, for \texttt{O1=[a,c]} and \texttt{O2=[b,c]}, 
    \texttt{c} is a shared reference.

\subsection{User Interface}
\label{sec:ui}

Our capture module works out of the box right after installation, requiring no code modification. Attaching our capture module (i.e., \capture) to a target application is as simple as switching the Python execution command \texttt{python target.py ...} to:
\begin{center}
\texttt{python \textbf{-m capture [args]} target.py ...}
\end{center}

\noindent
\capture is mindful of its overhead over regular application execution. 
Based on their use cases, users can adjust \capture's cost and quality such as overhead allowances on execution time, memory, and/or storage.
In the future, we aim to support fine-grained controls beyond the out-of-the-box capture, for example, object inclusion/exclusion, user-defined object serialization, capture location hint, and programmatic capture invocation.

\ignore{
\subsection{Internal Inspection Logic}
\label{sec:overview:inspection}

The core premise of this project is 
    the ability to have \emph{inspection points} as follows,
allowing our framework to monitor statements
    and examine the changes made to an application state:
\begin{lstlisting}[
    basicstyle=\ttfamily\small\linespread{1.0}\selectfont,
    xleftmargin=0.4cm,
    language=bash,
    morekeywords={preprocess,postprocess}
    ]
for stmt in statements do
    preprocess(stmt)    # our inspection
    execute(stmt)       # regular execution
    postprocess(stmt)   # our inspection
\end{lstlisting}
where \texttt{stmt} is part of a user program executed at a time
    appearing at the global level or inside a function definition.
    
Utilizing the existing Python kernel, at least three approaches
    allow non-intrusive inspection.
The first is to periodically interrupt the main program
    using \texttt{threading.Timer()}.
At the interrupted point, our \texttt{capture} gains control,
    allowing the inspection of the current program and data states.
The second approach is to observe every statement
    using \texttt{sys.settrace()},
        an approach used by the Python debugger (\texttt{pbd}).
Finally,
    IPython-based applications (e.g., Jupyter)
allows cell-level inspections via custom magic functions.
In this work,
    we take the first approach due to its low overhead.
}

\subsection{Efficient State Persistence Allows DART}
\label{sec:overview:dart}

In the example in \cref{sec:intro},
    Jane can ultimately obtain results by re-executing the code.
This work aims to avoid this time-consuming process
    by persisting a minimal amount of information.


\paragraph{Interpreter Generates Redo Log}

We can reconstruct a Python application state
    by re-executing its statements as occurred in the past.
In DBMS, durability is achieved
    via logs~\cite{mohan1992aries}.
That is,
    by replaying DML statements
        recorded in write-ahead logs,
    we can reach the desired state.
Likewise,
    the Python kernel and a program serve as a \emph{blueprint}
for generating logs (i.e., statements).
    By re-executing them until a target point,
        we can reconstruct an application state.


\ignore{
\paragraph{Re-execution is Slow}

Re-executing statements can be slow
in data-intensive applications
    since it may involve
    compute-/data-intensive ML training,
which we aim to avoid in achieving DART.
}

\ignore{
\paragraph{Persisted States Enable Instant Replay}

We can fast-forward executions if we know their outcome,
    which we can save onto durable storage
        for future reconstruction.
DBMS uses \emph{checkpoints}~\cite{arulraj2016write},
    the snapshots containing valid states of a database.
Upon failure,
    the latest state can be quickly reconstructed
        by replaying the logs on a checkpoint.
In Python, however,
    periodically creating checkpoints can incur significant overhead
        in large-scale machine learning and data analytics.
These challenges are discussed in depth in \cref{sec:overview:challenge}.
}


\paragraph{Instant Replay for DART}

Like DBMS checkpoints~\cite{arulraj2016write}, we can persist Python checkpoints and replay to an arbitrary state to enable DART.
\emph{Durability:}
We can survive failures
    by storing sufficient information (for instant replays)
        on durable storage.
\emph{Atomicity:}
Only succeeded statements can be replayed one at a time,
        preventing partial updates.
\emph{Replicability:}
We can copy the \emph{sufficient information}
        to a target machine for application resumption.
\emph{Time-Versioning:}
We can replay an application
    up until the moment of interest.

\subsection{Use Cases of DART}
\label{sec:overview:usecases}


DART enables several practical applications, immediately.

\begin{enumerate}
\item Model Checkpointing: 
During training, models are automatically saved.
In contrast to 
    checkpointing within ML frameworks~\cite{vartak2018mistique} or 
    manually exporting to storage~\cite{zaharia2018accelerating}, 
our method requires no explicit code changes.
Furthermore, capturing entire states 
    (not limited to specified objects)
allows easier analyses.
%
\item Time-Travel Diagnosis: 
With instant replay to a versioned state, 
we can inspect past states of objects,
    allowing easier diagnosis of learning trajectories,
    NaN errors, exploding gradients, etc.
Compared to existing tools~\cite{visan2011urdb},
    we incur low overhead.
%
\item Fault Recovery: 
ML pipelines may fail due to invalid data, 
    ill-conditioned tasks, modeling instability, or 
    insufficient resources. 
Upon failure, the user can recover from a persisted state.
Compared to restarting from scratch,
we offer faster recovery.
\end{enumerate}


\subsection{Technical Challenges in Persisting States}
\label{sec:overview:challenge}

Simply saving entire states
    (including all the objects such as
        training/test data, trained models)
can impose a significant overhead
    in regular program executions.
In developing our framework,
    we have encountered issues
        arising in three different dimensions:
data redundancy, object change identification, 
    and object inter-dependencies.

\paragraph{Data Redundancy}

Two different states $S_i$ and $S_j$ may include
    duplicate, unmodified objects.
This occurs frequently since a statement, e.g., \texttt{df=df.filter()}, may modify a subset of objects.
If we can efficiently identify
    whether an object has been modified,
we can avoid storing redundant data.

\paragraph{Identifying Modified Objects}

Identifying object modifications is technically challenging.
First, the declaration of constant or immutable variables is 
    not supported by the Python language;
    thus, every object is subject to change.
Second, unlike DBMS tables, 
    an object may consist of other objects,
    requiring recursive inspections.
Third,
    variable names may point to the same object.
For instance,
    \texttt{v1.update()}
    may cause a change to \texttt{v2} pointing to the same object.

\paragraph{Object Inter-Dependencies}

Persisting states (with objects)
    must be aware of their internal object references
for correctness.
Suppose two lists, \texttt{o1=[a,c]} and \texttt{o2=[b,c]},
    sharing \texttt{c}.
If we persist \texttt{o1} and \texttt{o2} separately 
    (for example, because they existed in two different states)
and load them separately,
    we have \texttt{o1=[a,c1]} and \texttt{o2=[b,c2]}
where \texttt{c1} and \texttt{c2} are two deep-copied objects 
    containing the same value as \texttt{c}.
That is,
    a reference-unaware method may
        unintentionally break the relationship 
        between \texttt{o1} and \texttt{o2}.

\subsection{Why Python?}
\label{sec:overview:python}

Although the concept of transactional systems is applicable to other languages, Python is currently a good starting point.
First, Python is a widely used language especially in ML applications; therefore, the solution to transactional Python with a minimal user effort will be impactful.
Secondly, thanks to its extensibility to lower-level languages, Python as a declarative language like SQL sets a data boundary and scale appropriate for transactions. 
For example, NumPy~\cite{harris2020numpy} programmers now manipulate their tensors without specifying details like vectorization and broadcasting.
Declarative Python alleviates transaction workloads by separating long-lived high-level objects in Python and short-lived low-level objects in lower-level languages. Moreover, these high-level objects are likely independent of the environment, increasing the replicability across platforms.
Lastly, Python implementations (e.g., CPython~\cite{cpython}) support building blocks necessary for our solution, including state extraction for capturing and dynamic execution for recovery. On the contrary, for example, C/C++ state extraction is much more difficult because its objects reside at arbitrary memory addresses.

\section{State Delta for Fast Persistence}
\label{sec:state}


    Persisting an entire state---encompassing models, datasets, and intermediates---can incur prohibitive costs.
To avoid them,
    we identify \emph{partial changes} between states---which we call \emph{deltas}.
In turn, this reduces data redundancy and saves persistence costs like CPU, I/O, and storage.
This section describes the workflow (\cref{sec:state-workflow}), 
    efficient approaches to identifying delta (\cref{sec:change:identify}), 
    and future directions (\cref{sec:change:dynamic}).

\subsection{Workflow}
\label{sec:state-workflow}

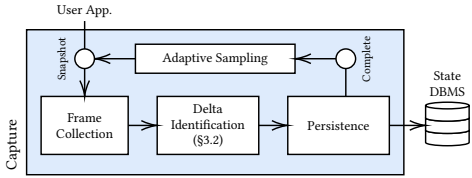
\begin{figure}[t]
  \centering
  \resizebox{0.75\linewidth}{!}{\tikzset{every picture/.style={line width=0.75pt}} 

\begin{tikzpicture}[x=0.75pt,y=0.75pt,yscale=-1,xscale=1]

\draw  [fill={rgb, 255:red, 222; green, 235; blue, 252 }  ,fill opacity=1 ] (20,20) -- (280,20) -- (280,120) -- (20,120) -- cycle ;
\draw  [fill={rgb, 255:red, 255; green, 255; blue, 255 }  ,fill opacity=1 ] (30,66) -- (90,66) -- (90,106) -- (30,106) -- cycle ;
\draw  [fill={rgb, 255:red, 255; green, 255; blue, 255 }  ,fill opacity=1 ] (110,66) -- (180,66) -- (180,106) -- (110,106) -- cycle ;
\draw  [fill={rgb, 255:red, 255; green, 255; blue, 255 }  ,fill opacity=1 ] (200,66) -- (270,66) -- (270,106) -- (200,106) -- cycle ;
\draw    (90,86) -- (108,86) ;
\draw [shift={(110,86)}, rotate = 180] [color={rgb, 255:red, 0; green, 0; blue, 0 }  ][line width=0.75]    (10.93,-3.29) .. controls (6.95,-1.4) and (3.31,-0.3) .. (0,0) .. controls (3.31,0.3) and (6.95,1.4) .. (10.93,3.29)   ;
\draw    (180,86) -- (198,86) ;
\draw [shift={(200,86)}, rotate = 180] [color={rgb, 255:red, 0; green, 0; blue, 0 }  ][line width=0.75]    (10.93,-3.29) .. controls (6.95,-1.4) and (3.31,-0.3) .. (0,0) .. controls (3.31,0.3) and (6.95,1.4) .. (10.93,3.29)   ;
\draw    (60,14) -- (60,64) ;
\draw [shift={(60,66)}, rotate = 270] [color={rgb, 255:red, 0; green, 0; blue, 0 }  ][line width=0.75]    (10.93,-3.29) .. controls (6.95,-1.4) and (3.31,-0.3) .. (0,0) .. controls (3.31,0.3) and (6.95,1.4) .. (10.93,3.29)   ;
\draw  [fill={rgb, 255:red, 255; green, 255; blue, 255 }  ,fill opacity=1 ] (325,89.24) -- (325,97.72) .. controls (325,98.98) and (318.28,100) .. (310,100) .. controls (301.72,100) and (295,98.98) .. (295,97.72) -- (295,89.24)(325,89.24) .. controls (325,90.5) and (318.28,91.52) .. (310,91.52) .. controls (301.72,91.52) and (295,90.5) .. (295,89.24) .. controls (295,87.98) and (301.72,86.96) .. (310,86.96) .. controls (318.28,86.96) and (325,87.98) .. (325,89.24) -- cycle ;
\draw  [fill={rgb, 255:red, 255; green, 255; blue, 255 }  ,fill opacity=1 ] (325,80.76) -- (325,89.24) .. controls (325,90.5) and (318.28,91.52) .. (310,91.52) .. controls (301.72,91.52) and (295,90.5) .. (295,89.24) -- (295,80.76)(325,80.76) .. controls (325,82.02) and (318.28,83.04) .. (310,83.04) .. controls (301.72,83.04) and (295,82.02) .. (295,80.76) .. controls (295,79.5) and (301.72,78.48) .. (310,78.48) .. controls (318.28,78.48) and (325,79.5) .. (325,80.76) -- cycle ;
\draw  [fill={rgb, 255:red, 255; green, 255; blue, 255 }  ,fill opacity=1 ] (325,72.28) -- (325,80.76) .. controls (325,82.02) and (318.28,83.04) .. (310,83.04) .. controls (301.72,83.04) and (295,82.02) .. (295,80.76) -- (295,72.28)(325,72.28) .. controls (325,73.54) and (318.28,74.57) .. (310,74.57) .. controls (301.72,74.57) and (295,73.54) .. (295,72.28) .. controls (295,71.02) and (301.72,70) .. (310,70) .. controls (318.28,70) and (325,71.02) .. (325,72.28) -- cycle ;
\draw  [fill={rgb, 255:red, 255; green, 255; blue, 255 }  ,fill opacity=1 ] (95,30) -- (205,30) -- (205,50) -- (95,50) -- cycle ;
\draw    (95,40) -- (68.75,40) ;
\draw [shift={(66.75,40)}, rotate = 360] [color={rgb, 255:red, 0; green, 0; blue, 0 }  ][line width=0.75]    (10.93,-3.29) .. controls (6.95,-1.4) and (3.31,-0.3) .. (0,0) .. controls (3.31,0.3) and (6.95,1.4) .. (10.93,3.29)   ;
\draw    (240,66) -- (240,40) -- (207,40) ;
\draw [shift={(205,40)}, rotate = 360] [color={rgb, 255:red, 0; green, 0; blue, 0 }  ][line width=0.75]    (10.93,-3.29) .. controls (6.95,-1.4) and (3.31,-0.3) .. (0,0) .. controls (3.31,0.3) and (6.95,1.4) .. (10.93,3.29)   ;
\draw    (270,86) -- (293,86) ;
\draw [shift={(295,86)}, rotate = 180] [color={rgb, 255:red, 0; green, 0; blue, 0 }  ][line width=0.75]    (10.93,-3.29) .. controls (6.95,-1.4) and (3.31,-0.3) .. (0,0) .. controls (3.31,0.3) and (6.95,1.4) .. (10.93,3.29)   ;
\draw  [fill={rgb, 255:red, 255; green, 255; blue, 255 }  ,fill opacity=1 ] (53.25,40) .. controls (53.25,36.27) and (56.27,33.25) .. (60,33.25) .. controls (63.73,33.25) and (66.75,36.27) .. (66.75,40) .. controls (66.75,43.73) and (63.73,46.75) .. (60,46.75) .. controls (56.27,46.75) and (53.25,43.73) .. (53.25,40) -- cycle ;
\draw  [fill={rgb, 255:red, 255; green, 255; blue, 255 }  ,fill opacity=1 ] (233.25,40) .. controls (233.25,36.27) and (236.27,33.25) .. (240,33.25) .. controls (243.73,33.25) and (246.75,36.27) .. (246.75,40) .. controls (246.75,43.73) and (243.73,46.75) .. (240,46.75) .. controls (236.27,46.75) and (233.25,43.73) .. (233.25,40) -- cycle ;

\draw (60,86) node  [font=\footnotesize] [align=left] {\begin{minipage}[lt]{38.56pt}\setlength\topsep{0pt}
\begin{center}
Frame\\Collection
\end{center}

\end{minipage}};
\draw (145,86) node  [font=\footnotesize] [align=left] {\begin{minipage}[lt]{47.18pt}\setlength\topsep{0pt}
\begin{center}
Delta Identification (\cref{sec:change:identify})
\end{center}

\end{minipage}};
\draw (235,86) node  [font=\footnotesize] [align=left] {Persistence};
\draw (60,13) node [anchor=south] [inner sep=0.75pt]  [font=\footnotesize] [align=left] {User App.};
\draw (17,118) node [anchor=south west] [inner sep=0.75pt]  [font=\small,rotate=-270] [align=left] {Capture};
\draw (50.25,40) node [anchor=south] [inner sep=0.75pt]  [font=\scriptsize,rotate=-270] [align=left] {Snapshot};
\draw (150,40) node  [font=\footnotesize] [align=left] {Adaptive Sampling};
\draw (310,67) node [anchor=south] [inner sep=0.75pt]  [font=\footnotesize] [align=left] {\begin{minipage}[lt]{26.3pt}\setlength\topsep{0pt}
\begin{center}
State\\DBMS
\end{center}

\end{minipage}};
\draw (249.75,40) node [anchor=north] [inner sep=0.75pt]  [font=\scriptsize,rotate=-270] [align=left] {Complete};

\end{tikzpicture}}
  \vspace{-2mm}
  \caption{Component diagram inside our capture module.}
  \vspace{-3mm}
  \label{fig:capture_sketch}
\end{figure}

\cref{fig:capture_sketch} illustrates components inside \capture (\cref{sec:ui}). 
Initially, 
\capture sets up the necessary instruments and executes the target application as usual. Afterward, \capture periodically takes \textit{state snapshots} by collecting CPython frames, then finding and persisting differences between consecutive state snapshots.

\paragraph{Interpreter Support} 
Among many~\cite{pypy,jython,ironpython}, we rely on the standard CPython interpreter~\cite{cpython},
    which allows:
(1) accessing its state as a stack of \emph{frames}, where
each frame corresponds to a function scope and consists of in-scope objects, 
    source code, instructions, etc. 
(2) timer-based triggers with a custom signal handler that is
    invoked between Python statements, outside the underlying C code.

\paragraph{Frame Collection} Frame collection first accesses the current frame and walks the frame stack to extract those frames belonging to the target application. 
This step is computationally cheap, incurring only Python references without data copying.

\paragraph{Persistence} \capture then persists those write and delete \emph{deltas} to a DBMS, ready for recovery. 
The DBMS may independently checkpoint the delta to speed up the recovery process.

\paragraph{Adaptive Sampling} \capture can adapt the sampling frequency accordingly 
to the overhead allowance, past performance, and application demands. 
For example, if the application quickly creates larger deltas or the DBMS exerts backpressure, 
the capture module may reduce the sampling frequency to match the overhead goal.

\paragraph{Robustness} 
We design \capture to be failsafe: a failure within the module does not crash the target application. 
Expected failures include unsupported object serializations and persistence failures. 
This implies that some snapshots may be missing, which can be remedied by re-execution during recovery. 
The snapshot after the failure needs to either cover deltas 
with respect to the snapshot before the failure or record a complete state.

\subsection{Correct and Efficient Delta Identification}
\label{sec:change:identify}

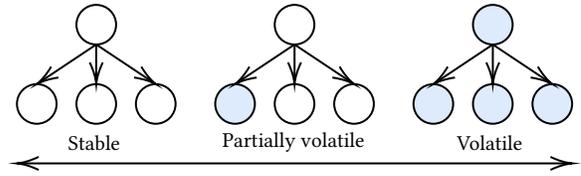
\begin{figure}[t]
  \centering
  \tikzset{every picture/.style={line width=0.75pt}} 

\begin{tikzpicture}[x=0.75pt,y=0.75pt,yscale=-1,xscale=1]

\draw   (40,20) .. controls (40,14.48) and (44.48,10) .. (50,10) .. controls (55.52,10) and (60,14.48) .. (60,20) .. controls (60,25.52) and (55.52,30) .. (50,30) .. controls (44.48,30) and (40,25.52) .. (40,20) -- cycle ;
\draw   (10,60) .. controls (10,54.48) and (14.48,50) .. (20,50) .. controls (25.52,50) and (30,54.48) .. (30,60) .. controls (30,65.52) and (25.52,70) .. (20,70) .. controls (14.48,70) and (10,65.52) .. (10,60) -- cycle ;
\draw   (40,60) .. controls (40,54.48) and (44.48,50) .. (50,50) .. controls (55.52,50) and (60,54.48) .. (60,60) .. controls (60,65.52) and (55.52,70) .. (50,70) .. controls (44.48,70) and (40,65.52) .. (40,60) -- cycle ;
\draw   (70,60) .. controls (70,54.48) and (74.48,50) .. (80,50) .. controls (85.52,50) and (90,54.48) .. (90,60) .. controls (90,65.52) and (85.52,70) .. (80,70) .. controls (74.48,70) and (70,65.52) .. (70,60) -- cycle ;
\draw    (50,30) -- (21.66,48.89) ;
\draw [shift={(20,50)}, rotate = 326.31] [color={rgb, 255:red, 0; green, 0; blue, 0 }  ][line width=0.75]    (10.93,-3.29) .. controls (6.95,-1.4) and (3.31,-0.3) .. (0,0) .. controls (3.31,0.3) and (6.95,1.4) .. (10.93,3.29)   ;
\draw    (50,30) -- (50,48) ;
\draw [shift={(50,50)}, rotate = 270] [color={rgb, 255:red, 0; green, 0; blue, 0 }  ][line width=0.75]    (10.93,-3.29) .. controls (6.95,-1.4) and (3.31,-0.3) .. (0,0) .. controls (3.31,0.3) and (6.95,1.4) .. (10.93,3.29)   ;
\draw    (50,30) -- (78.34,48.89) ;
\draw [shift={(80,50)}, rotate = 213.69] [color={rgb, 255:red, 0; green, 0; blue, 0 }  ][line width=0.75]    (10.93,-3.29) .. controls (6.95,-1.4) and (3.31,-0.3) .. (0,0) .. controls (3.31,0.3) and (6.95,1.4) .. (10.93,3.29)   ;
\draw   (140,20) .. controls (140,14.48) and (144.48,10) .. (150,10) .. controls (155.52,10) and (160,14.48) .. (160,20) .. controls (160,25.52) and (155.52,30) .. (150,30) .. controls (144.48,30) and (140,25.52) .. (140,20) -- cycle ;
\draw  [fill={rgb, 255:red, 222; green, 235; blue, 252 }  ,fill opacity=1 ] (110,60) .. controls (110,54.48) and (114.48,50) .. (120,50) .. controls (125.52,50) and (130,54.48) .. (130,60) .. controls (130,65.52) and (125.52,70) .. (120,70) .. controls (114.48,70) and (110,65.52) .. (110,60) -- cycle ;
\draw   (140,60) .. controls (140,54.48) and (144.48,50) .. (150,50) .. controls (155.52,50) and (160,54.48) .. (160,60) .. controls (160,65.52) and (155.52,70) .. (150,70) .. controls (144.48,70) and (140,65.52) .. (140,60) -- cycle ;
\draw  [fill={rgb, 255:red, 0; green, 0; blue, 0 }  ,fill opacity=0 ] (170,60) .. controls (170,54.48) and (174.48,50) .. (180,50) .. controls (185.52,50) and (190,54.48) .. (190,60) .. controls (190,65.52) and (185.52,70) .. (180,70) .. controls (174.48,70) and (170,65.52) .. (170,60) -- cycle ;
\draw    (150,30) -- (121.66,48.89) ;
\draw [shift={(120,50)}, rotate = 326.31] [color={rgb, 255:red, 0; green, 0; blue, 0 }  ][line width=0.75]    (10.93,-3.29) .. controls (6.95,-1.4) and (3.31,-0.3) .. (0,0) .. controls (3.31,0.3) and (6.95,1.4) .. (10.93,3.29)   ;
\draw    (150,30) -- (150,48) ;
\draw [shift={(150,50)}, rotate = 270] [color={rgb, 255:red, 0; green, 0; blue, 0 }  ][line width=0.75]    (10.93,-3.29) .. controls (6.95,-1.4) and (3.31,-0.3) .. (0,0) .. controls (3.31,0.3) and (6.95,1.4) .. (10.93,3.29)   ;
\draw    (150,30) -- (178.34,48.89) ;
\draw [shift={(180,50)}, rotate = 213.69] [color={rgb, 255:red, 0; green, 0; blue, 0 }  ][line width=0.75]    (10.93,-3.29) .. controls (6.95,-1.4) and (3.31,-0.3) .. (0,0) .. controls (3.31,0.3) and (6.95,1.4) .. (10.93,3.29)   ;
\draw  [fill={rgb, 255:red, 222; green, 235; blue, 252 }  ,fill opacity=1 ] (240,20) .. controls (240,14.48) and (244.48,10) .. (250,10) .. controls (255.52,10) and (260,14.48) .. (260,20) .. controls (260,25.52) and (255.52,30) .. (250,30) .. controls (244.48,30) and (240,25.52) .. (240,20) -- cycle ;
\draw  [fill={rgb, 255:red, 222; green, 235; blue, 252 }  ,fill opacity=1 ] (210,60) .. controls (210,54.48) and (214.48,50) .. (220,50) .. controls (225.52,50) and (230,54.48) .. (230,60) .. controls (230,65.52) and (225.52,70) .. (220,70) .. controls (214.48,70) and (210,65.52) .. (210,60) -- cycle ;
\draw  [fill={rgb, 255:red, 222; green, 235; blue, 252 }  ,fill opacity=1 ] (240,60) .. controls (240,54.48) and (244.48,50) .. (250,50) .. controls (255.52,50) and (260,54.48) .. (260,60) .. controls (260,65.52) and (255.52,70) .. (250,70) .. controls (244.48,70) and (240,65.52) .. (240,60) -- cycle ;
\draw  [fill={rgb, 255:red, 222; green, 235; blue, 252 }  ,fill opacity=1 ] (270,60) .. controls (270,54.48) and (274.48,50) .. (280,50) .. controls (285.52,50) and (290,54.48) .. (290,60) .. controls (290,65.52) and (285.52,70) .. (280,70) .. controls (274.48,70) and (270,65.52) .. (270,60) -- cycle ;
\draw    (250,30) -- (221.66,48.89) ;
\draw [shift={(220,50)}, rotate = 326.31] [color={rgb, 255:red, 0; green, 0; blue, 0 }  ][line width=0.75]    (10.93,-3.29) .. controls (6.95,-1.4) and (3.31,-0.3) .. (0,0) .. controls (3.31,0.3) and (6.95,1.4) .. (10.93,3.29)   ;
\draw    (250,30) -- (250,48) ;
\draw [shift={(250,50)}, rotate = 270] [color={rgb, 255:red, 0; green, 0; blue, 0 }  ][line width=0.75]    (10.93,-3.29) .. controls (6.95,-1.4) and (3.31,-0.3) .. (0,0) .. controls (3.31,0.3) and (6.95,1.4) .. (10.93,3.29)   ;
\draw    (250,30) -- (278.34,48.89) ;
\draw [shift={(280,50)}, rotate = 213.69] [color={rgb, 255:red, 0; green, 0; blue, 0 }  ][line width=0.75]    (10.93,-3.29) .. controls (6.95,-1.4) and (3.31,-0.3) .. (0,0) .. controls (3.31,0.3) and (6.95,1.4) .. (10.93,3.29)   ;
\draw    (12,90) -- (288,90) ;
\draw [shift={(290,90)}, rotate = 180] [color={rgb, 255:red, 0; green, 0; blue, 0 }  ][line width=0.75]    (10.93,-3.29) .. controls (6.95,-1.4) and (3.31,-0.3) .. (0,0) .. controls (3.31,0.3) and (6.95,1.4) .. (10.93,3.29)   ;
\draw [shift={(10,90)}, rotate = 0] [color={rgb, 255:red, 0; green, 0; blue, 0 }  ][line width=0.75]    (10.93,-3.29) .. controls (6.95,-1.4) and (3.31,-0.3) .. (0,0) .. controls (3.31,0.3) and (6.95,1.4) .. (10.93,3.29)   ;

\draw (49,79.5) node  [font=\small] [align=left] {Stable};
\draw (149.5,79.5) node  [font=\small] [align=left] {Partially volatile};
\draw (248.5,79.5) node  [font=\small] [align=left] {Volatile};

\end{tikzpicture}
  \vspace{-1.5mm}
  \caption{Spectrum of object volatility.}
  \vspace{-3mm}
  \label{fig:obj_volatility}
\end{figure}

There are two immediate approaches to identify state delta.

\paragraph{Approach 1: Per-variable Serialization} 

Apart from instructions and execution variables, the extracted frames also contain all objects in a target application, 
    including the global and local variables. 
Using Dill~\cite{dumpsession}, \capture serializes these objects separately. 
That is, it turns $n$ objects into $n$ byte arrays. Note that objects and nested objects with shared references are each serialized only once in one of these byte arrays. \capture then finds the difference between the previous snapshot's byte arrays and the current ones. If a byte array is new or changed, the capture module signals an (over)write delta. 
If one is missing, it signals a deletion delta.

\paragraph{Approach 2: ID Graph}
As an alternative to serialization, \capture also expresses Python objects by constructing an \textit{ID graph}.
Each object is assigned a unique ID~\cite{pythonid}
    based on its memory address during its lifetime. 
The ID graph contains $n$ nodes for each of the $n$ objects, and a directed edge $(n, m)$ exists in the ID graph if the object with ID $m$ is reachable from the object with ID $n$ via a reference.
Differences between snapshots can be similarly identified using the ID graph: new and modified nodes signal (over)write deltas, and missing nodes signal deletion deltas.


\paragraph{Pros and Cons of Two Approaches} The two approaches prefer different levels of object volatility. Per-variable serialization performs better towards the extremes (left and right ends of \cref{fig:obj_volatility}), where objects either remain the same or change entirely. This is because serialization on objects as a whole promotes defragmentation, thus more compact format.
On the other hand, the ID graph performs better in the middle of \cref{fig:obj_volatility}, 
    where only some part(s) of objects change.
Here state delta captures the differences more precisely, 
    so it would later reduce the size of serialized byte arrays and subsequently persistence costs.

When some objects or parts thereof implement equality operators (\texttt{\_\_eq\_\_}), 
the ID graph could be significantly faster than per-variable serialization. 
The difference is more pronounced if large objects implement their operators and the operators are efficient, for example, via mutable borrow checking, dirty bits, or intrinsic properties. 
However, equality operators are likely unavailable especially in higher-level objects while serialization is required to successfully capture a snapshot. Serialization is therefore more general and suitable as a fallback for both methods.

\subsection{Ongoing Effort toward Dynamic ID Graph}
\label{sec:change:dynamic}

Currently, we are exploring solutions to reduce the delta size by adaptively finding differences in smaller parts of objects.
Apart from object's volatility (\cref{fig:obj_volatility}), our delta identification should also consider object's decomposability. 
Highly decomposable objects are those having equally sized parts so that we can effectively separate portions of the delta. 
Together, decomposable and partially volatile objects are the ideal candidates for the ID graph approach to dynamically find sizable-and-volatile object parts and spare the other sizable-and-stable object parts. 
This dynamic ID graph should identify these candidates during execution with low overhead.



\begin{figure}[t]
\centering

\def\height{30mm}
\def\subfigurewidthA{\linewidth}
\def\plotwidth{85mm}
\def\subfigurevlabelgap{-2mm}
\def\barwidth{3mm}

\pgfplotstableread{
app	control	tens	tensOver	alltens	alltensOver	tensOverP	alltensOverP	tensfix	tensfixOver	alltensfix	alltensfixOver	tensfixOverP	alltensfixOverP	relOver
3	242.2240	246.8600	+4.6	253.5567	+11.3	1.9139	4.6786	246.9	+4.6	253.6	+11.3	1.9139	4.6786	2.4445
4	619.0600	715.4640	+96.4	740.4700	+121.4	15.5726	19.6120	715.5	+96.4	740.5	+121.4	15.5726	19.6120	1.2594
1	47.4030	62.0720	+14.7	69.1467	+21.7	30.9453	45.8698	48.1	+0.7	51.0	+3.6	1.4774	7.4967	5.0743
2	234.3700	363.5290	+129.2	454.1100	+219.7	55.1090	93.7577	258.9	+24.6	302.4	+68.1	10.4749	29.0467	2.7730
} \overheadTime

\begin{tikzpicture}  
\begin{axis}[
    width=\plotwidth,
    height=\height,
    axis lines=left,
    ymin=0.0,
    ymax=35,
    ylabel={Overhead (\%)},
    ytick={0,10,20,30},
    minor y tick num=1,
    minor grid style=lightgray,
    ymajorgrids,
    yminorgrids,
    enlarge x limits=0.15,
    xtick={1, 2, 3, 4},
    xticklabels={\texttt{skl\_kmeans}, \texttt{skl\_tsne}, \texttt{pytorch\_mnist}, \texttt{pytorch\_dcgan}},
    ybar=1pt,
    bar width=\barwidth,
    area legend,
    xlabel near ticks,
    nodes near coords,
    nodes near coords align={vertical},
    nodes near coords style={
        anchor=east,
        rotate=25,
        font=\scriptsize,
        right,
        #1
    },
    legend image code/.code={%
        \draw[#1, draw=none] (0cm,-0.1cm) rectangle (0.6cm,0.1cm);
    },
    xticklabel style={rotate=0},
    legend style={
        nodes={scale=0.7, transform shape}, 
        at={(0.5,1.05)},
        anchor=south,
        /tikz/every even column/.append style={column sep=3mm},
        draw=black,
        font=\footnotesize\bfseries},
    legend cell align={center},
    legend columns=-1,
    every axis/.append style={font=\footnotesize},
]
    
    \addplot[fill={Dcolor}, point meta=explicit symbolic]
    plot [error bars/.cd, y dir=plus,y explicit,error bar style={color=black}]
    table[x=app,y=alltensfixOverP,meta=alltensfixOver] {\overheadTime};
    \addlegendentry{Captured Whole State};
    
    \addplot[fill={Bcolor}, point meta=explicit symbolic]
    plot [error bars/.cd, y dir=plus,y explicit,error bar style={color=black}]
    table[x=app,y=tensfixOverP,meta=tensfixOver] {\overheadTime};
    \addlegendentry{Ours (Delta)};
        
\end{axis}
\end{tikzpicture}

\vspace{-3mm}
\caption{Execution time overheads (\%) over execution times without capture. Numbers denote absolute overheads (sec).}
\label{fig:ml_overhead_time}
\vspace{-3.5mm}
\end{figure}
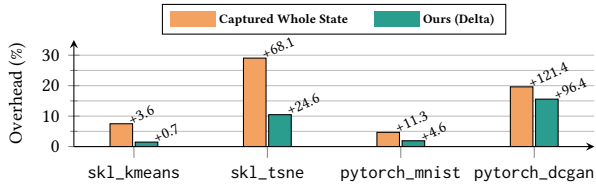

\section{Feasibility Study}
\label{sec:exp}

We implement a proof-of-concept capture module in Python. This implementation follows \cref{fig:capture_sketch} with per-variable serialization, persistence onto local disks, and fixed persistence frequency. 
Our empirical study shows that the module has a low execution time overhead (\cref{sec:exp-time}) and its storage cost grows at an expected pace (\cref{sec:exp-storage}).

\subsection{Evaluation Setup}
\label{sec:exp-setup}

\paragraph{Workloads} Our workloads consist of four ML applications. 
The first two are from the scikit-learn benchmark repository~\cite{sklbenchmark}. 
The other two are official PyTorch examples~\cite{pytorchexample}. 
We adopt the same configurations given in those sources. 

\begin{enumerate}
\item \textbf{\texttt{skl\_kmeans}}: K-means from scikit-learn finds clusters among 1M samples drawn from 1000 isotropic Gaussian blobs with 20 feature dimensions~\cite{sklbenchmarkKmeans}.
\item \textbf{\texttt{skl\_tsne}}: t-SNE from scikit-learn visualizes high-dimensional images 
    after embedding them in a low-dimensional space~\cite{sklbenchmarkTsne}.
\item \textbf{\texttt{pytorch\_mnist}}: PyTorch's 2-layer convolutional neural network (CNN) with 2 fully connected layers trains to classify MNIST images into digits~\cite{pytorchexampleMnist}.
\item \textbf{\texttt{pytorch\_dcgan}}: PyTorch's deep convolutional generative adversarial network (DCGAN) trains to generate CIFAR-10 images by balancing an adversarial pair of 5-layers CNNs~\cite{pytorchexampleDcgan}.
\end{enumerate}

\paragraph{System Environments} 
All experiments run on MacBook Pro 2020 with M1, 16 GB memory, and SSD storage under APFS. 
We set the sampling frequency to capture a snapshot every 10 seconds.

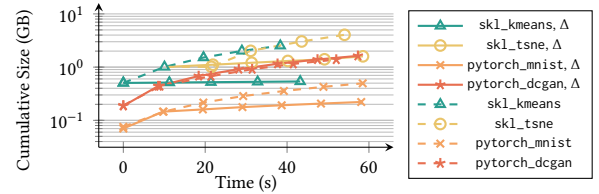
\begin{figure}[t]
\centering

\def\height{35mm}
\def\subfigurewidthA{\linewidth}
\def\plotwidth{55mm}
\def\subfigurevlabelgap{-2mm}
\def\barwidth{3mm}

\pgfplotstableread{
time cdelta
0.0 0.070078485
9.697670221328735 0.145694863
19.431233167648315 0.1603221
29.09994912147522 0.177308828
38.68781018257141 0.191936065
48.28802037239075 0.206563223
58.11606407165527 0.221190381
} \cdeltatensmnist
\pgfplotstableread{
time cdelta
0.0 0.186525336
8.504585981369019 0.439534509
18.40294075012207 0.67995994
28.211817979812622 0.91619078
37.67475485801697 1.154519042
47.43556189537048 1.390749858
57.30648493766785 1.626980573
} \cdeltatensdcgan

\pgfplotstableread{
time cdelta
0.0 0.0
10.882973909378052 1.011812851
21.880337953567505 1.109661998
31.35170602798462 1.207511145
40.190922021865845 1.305360292
49.14220595359802 1.403209439
58.540647983551025 1.5953067
} \cdeltatenssktsne  

\pgfplotstableread{
time cdelta
0.0 0.504525997
11.249054908752441 0.512866489
21.405282974243164 0.521206947
32.808435916900635 0.529547405
43.20724081993103 0.537887863
} \cdeltatensskkmeans  
\pgfplotstableread{
time cdelta
0.0 0.075616469
9.829596996307373 0.145694994
19.55088710784912 0.215773519
29.28644299507141 0.285852044
39.2232460975647 0.355930569
48.88498401641846 0.426009094
58.480602979660034 0.496087619
} \cdeltaalltensmnist
\pgfplotstableread{
time cdelta
0.0 0.19071964
8.93604588508606 0.438486452
21.39554190635681 0.674718237
31.46980309486389 0.913047498
41.51013994216919 1.149279283
52.02086591720581 1.385511068
} \cdeltaalltensdcgan

\pgfplotstableread{
time cdelta
0.0 0.504525997
10.05414605140686 1.009052064
19.588874101638794 1.513578131
28.969818830490112 2.018104198
38.345184087753296 2.522630265
} \cdeltaalltensskkmeans  

\pgfplotstableread{
time cdelta
0.0 0.0
10.424057006835938 1.011812851
21.52614402770996 1.011812851
31.114086151123047 2.023625778
43.57103681564331 3.035438705
54.03727102279663 4.047251632
} \cdeltaalltenssktsne  

\begin{tikzpicture}
\begin{axis}[
    width=\plotwidth,
    height=\height,
    axis lines=left,
    log origin=infty,
    ymode=log,
    ymin=0.03125,
    ymax=16,
    ylabel={Cumulative Size (GB)},
    minor y tick num=1,
    minor grid style=lightgray,
    ymajorgrids,
    yminorgrids,
    xmin=0,
    xmax=60,
    enlarge x limits=0.1,
    xticklabel style={rotate=0},
    xlabel={Time (s)},
    xlabel near ticks,
    xlabel shift=-2mm,
    y label style={xshift=-2mm},
    legend style={
        nodes={scale=0.7, transform shape}, 
        at={(1.05,0.95)},
        anchor=north west,
        draw=black,
        font=\small},
    legend cell align={center},
    legend columns=1,
    every axis/.append style={font=\footnotesize}
]
    \addplot[mark=triangle, color={Bcolor}, thick, mark repeat=1, mark phase=0]
    table[x=time,y=cdelta] {\cdeltatensskkmeans};
    \addlegendentry{\texttt{skl\_kmeans}, $\Delta$};
    
    \addplot[mark=o, color={Ccolor}, thick, mark repeat=1, mark phase=2]
    table[x=time,y=cdelta] {\cdeltatenssktsne};
    \addlegendentry{\texttt{skl\_tsne}, $\Delta$};
    
    \addplot[mark=x, color={Dcolor}, thick, mark repeat=1]
    table[x=time,y=cdelta] {\cdeltatensmnist};
    \addlegendentry{\texttt{pytorch\_mnist}, $\Delta$};
    
    \addplot[mark=star, color={Ecolor}, thick, mark repeat=1, mark phase=0]
    table[x=time,y=cdelta] {\cdeltatensdcgan};
    \addlegendentry{\texttt{pytorch\_dcgan}, $\Delta$};

    \addplot[mark=triangle, dashed, color={Bcolor}, thick, mark repeat=1, mark phase=0,mark options={solid,fill=Bcolor}]
    table[x=time,y=cdelta] {\cdeltaalltensskkmeans};
    \addlegendentry{\texttt{skl\_kmeans}};
    
    \addplot[mark=o, dashed, color={Ccolor}, thick, mark repeat=1, mark phase=2,mark options={solid,fill=Ccolor}]
    table[x=time,y=cdelta] {\cdeltaalltenssktsne};
    \addlegendentry{\texttt{skl\_tsne}};
    
    \addplot[mark=x, dashed, color={Dcolor}, thick, mark repeat=1,mark options={solid,fill=Dcolor}]
    table[x=time,y=cdelta] {\cdeltaalltensmnist};
    \addlegendentry{\texttt{pytorch\_mnist}};
    
    \addplot[mark=star, dashed, color={Ecolor}, thick, mark repeat=1, mark phase=0,mark options={solid,fill=Ecolor}]
    table[x=time,y=cdelta] {\cdeltaalltensdcgan};
    \addlegendentry{\texttt{pytorch\_dcgan}};

\end{axis}
\end{tikzpicture}

\vspace{-5mm}
\caption{Storage cost in the first minute. Lines with ``\texttt{workload}, $\Delta$'' persists deltas; other lines persist whole states.}
\label{fig:ml_overhead_storage}
\vspace{-3.5mm}
\end{figure}



\subsection{Execution Time Overhead is Low}
\label{sec:exp-time}

At one snapshot every 10 seconds, captured ML applications run with moderate execution time overhead compared to those without capturing (\cref{fig:ml_overhead_time}). The relative overheads in order of the workloads are 1.5\%, 10.5\%, 1.9\%, and 15.6\% as opposed to 7.5\%, 29.0\%, 4.7\%, and 19.6\% overhead capturing without state delta. Higher overheads is due to larger sizes of state deltas. In \texttt{skl\_tsne} with a high relative overhead, the embedding model internally stores dataset samples and moves them around between training iterations; thus, our per-variable serialization detects delta at every snapshot. In addition, \texttt{pytorch\_dcgan} continuously trains its deep CNN discriminator model, incurring 180-240 MB every snapshot. ID graph approach could detect the shuffle and reduce this redundancy if the models reference samples instead of duplicating them. Alternatively, with a less frequent sampling, these overheads would decrease.

\subsection{Storage Cost is Affordable}
\label{sec:exp-storage}

\cref{fig:ml_overhead_storage} shows that the trajectories of storage usage with state delta are growing at much lower rates on almost all workloads. The only exception is \texttt{pytorch\_dcgan} where the model changes are dominant. Commonly across applications, initial delta snapshots are larger because of the static datasets. Afterward, the storage trends increase proportionally to the model sizes due to their updates. In order of the workloads, average snapshot sizes are 107.6, 146.2, 24.0, and 238.2 MB. As the most skewed workload, \texttt{skl\_kmeans} has a 504 MB initial snapshot followed by 8 MB snapshots.

\section{Conclusion}
\label{sec:conclusion}

This work proposes a novel approach, \emph{Transactional Python},
to bring durability, atomicity, replicability, and time-versioning (DART)
    to a wide range of data-intensive applications for
    machine learning and scientific computing.
Our proposed system design and 
    \emph{delta}-based efficient state persistence
    show the potential for providing DART
    with low computational overhead as well as affordable storage costs.
We will continue in this direction
    to develop dynamic ID graph construction
        and to test with a larger number of real-world applications
    in collaboration with industry partners.
    We also plan to accelerate delta searches for recovery by novel indexing techniques~\cite{chockchowwat2022airindex,chockchowwat2022airphant}.







\begin{acks}
This work is supported in part by
    the National Center for Supercomputing Applications
        and Microsoft Azure.
\end{acks}

\newpage
\bibliographystyle{ACM-Reference-Format}
\bibliography{_refs}


\end{document}